\documentclass[apj,iop]{emulateapj}

\usepackage{natbib}
\usepackage{longtable}
\usepackage{hyperref}

\shorttitle{X-rays from the Massive Black Hole in Henize 2-10}

\shortauthors{Reines et al.}

\begin{document}

\title{Deep Chandra Observations of the Compact Starburst Galaxy Henize 2-10: \\ X-rays from the Massive Black Hole}

\author{Amy E. Reines\altaffilmark{1}}
\affil{National Optical Astronomy Observatory, 950 North Cherry Avenue, Tucson, AZ 85719, USA}
\affil{Department of Astronomy, University of Michigan, 1085 South University Avenue, Ann Arbor, MI 48109, USA}
\email{reines@noao.edu}

\author{Mark T. Reynolds}
\affil{Department of Astronomy, University of Michigan, 1085 South University Avenue, Ann Arbor, MI 48109, USA}

\author{Jon M. Miller}
\affil{Department of Astronomy, University of Michigan, 1085 South University Avenue, Ann Arbor, MI 48109, USA}

\author{Gregory R. Sivakoff}
\affil{Department of Physics, University of Alberta, CCIS 4-181, Edmonton AB T6G 2E1, Canada}

\author{Jenny E. Greene}
\affil{Department of Astrophysical Sciences, Princeton University, Princeton, NJ 08544, USA}

\author{Ryan C. Hickox}
\affil{Department of Physics and Astronomy, Dartmouth College, 6127 Wilder Laboratory, Hanover, NH 03755, USA}

\and

\author{Kelsey E. Johnson}
\affil{Department of Astronomy, University of Virginia, P.O. Box 400325, Charlottesville, VA 22904-4325, USA}

\altaffiltext{1}{Hubble Fellow}

\begin{abstract}

We present follow-up X-ray observations of the candidate massive black hole (BH) in the nucleus of the low-mass, compact starburst galaxy Henize 2-10.  Using new high-resolution observations from the {\it Chandra X-ray Observatory} totaling 200~ks in duration, as well as archival {\it Chandra} observations from 2001, we demonstrate the presence of a previously unidentified X-ray point source that is spatially coincident with the known nuclear radio source in Henize 2-10 (i.e., the massive BH).  We show that the hard X-ray emission previously identified in the 2001 observation is dominated by a source that is distinct from the nucleus, with the properties expected for a high-mass X-ray binary.  The X-ray luminosity of the nuclear source suggests the massive BH is radiating significantly below its Eddington limit ($\sim$10$^{-6}~L_{\rm Edd}$), and the soft spectrum resembles other weakly accreting massive BHs including Sagittarius A$^{*}$.  Analysis of the X-ray light curve of the nucleus reveals the tentative detection of a $\sim 9$-hour periodicity, although additional observations are required to confirm this result.  Our study highlights the need for sensitive high-resolution X-ray observations to probe low-level accretion, which is the dominant mode of BH activity throughout the Universe.

\end{abstract}


\section{Introduction}\label{sec:intro}

\begin{figure*}[!t]
\begin{center}
\includegraphics[width=6.2in]{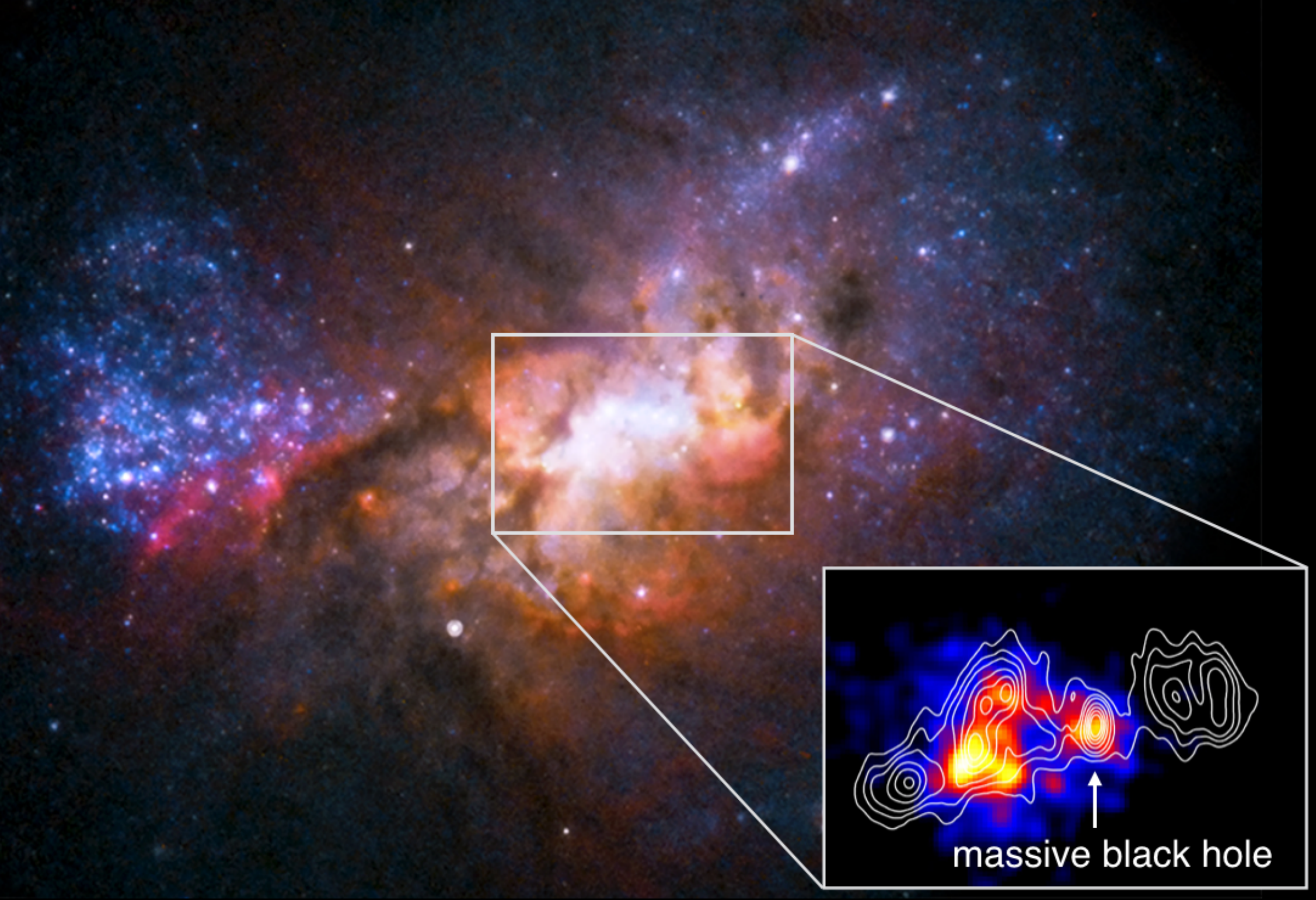}
\end{center}
\caption{{\it HST} image of Henize 2-10.  The inset shows our new 160 ks {\it Chandra} observation with VLA radio contours from \citet{reinesetal2011} and has dimensions $6\arcsec \times 4\arcsec$ ($\sim$ 265 pc $\times$ 175 pc).
\label{fig:he210}}
\end{figure*}

Henize 2-10 is a remarkable compact starburst galaxy, hosting an abundance of young ``super star clusters" \citep[e.g.,][]{johnsonetal2000} and a candidate low-luminosity active galactic nucleus (AGN) \citep{reinesetal2011}. 
The discovery of an AGN in Henize 2-10
provides an excellent opportunity to study BH accretion and star formation in a nearby ($\sim
9$ Mpc), low-mass \citep[$\lesssim 10^{10} M_\odot$;][]{nguyenetal2014}, gas-rich galaxy, as well as
the potential formation of a nuclear star cluster around a preexisting massive BH \citep{nguyenetal2014,arcaseddaetal2015}.  
Moreover, this finding has helped spark a number of recent searches for AGNs in dwarf galaxies \citep[e.g.,][]{reinesetal2013,reinesetal2014,baldassareetal2015,lemonsetal2015,baldassareetal2016,
hainlineetal2016}, ultimately leading to the realization that massive BHs in dwarfs
are much more common than previously thought (for a review, see \citealt{reinescomastri2016}).

The evidence for a massive BH in Henize~2-10 comes from a wealth of multi-wavelength data \citep{reinesetal2011},
including Very Large Array (VLA) radio observations that
reveal an unresolved non-thermal
nuclear point source (also see \citealt{kobulnickyjohnson1999} and \citealt{johnsonkobulnicky2003}).  Very long baseline
interferometry (VLBI) observations constrain the size of the nuclear radio emission to $< 1$ pc
$\times$ 3 pc and the high brightness temperature of the radio core confirms a non-thermal origin \citep{reinesdeller2012}.  

{\it Chandra X-ray Observatory} observations of Henize 2-10 taken in 2001 show point-like
hard X-ray emission that has previously been associated with the nuclear radio source
\citep{ottetal2005,kobulnickymartin2010,reinesetal2011}.  
Here we demonstrate that this emission is dominated by a source that is highly variable (also see \citealt{whalenetal2015}) and {\it not} in fact co-spatial with the radio source.  
Our new deep {\it Chandra} observations expose a different, previously unidentified X-ray counterpart to the nuclear radio source (Figure \ref{fig:he210}), for which we examine the X-ray spectrum and light curve.  

\section{Observations and Data Reduction}\label{sec:data}

We obtained new {\it Chandra} observations of Henize 2-10 in February 2015.  The total exposure time of $\sim$200~ks was broken up into two observations of 159066 s and 37577 s beginning on February 5 and 16 (PI:
Reines, ObsIDs 16068 and 16069). These observations were taken in VFAINT mode.  We also
retrieved the archival observation taken on 2001 March 23
(PI: Martin, ObsID 2075), which was taken in FAINT mode with
an exposure time of 19755 s.
In all three observations, the galaxy was placed on the S3 chip of the Advanced CCD
Imaging Spectrometer (ACIS) detector. The data were reduced and reprocessed with {\small
  CIAO} version 4.6 \citep{fruscioneetal2006} utilizing CALDB version 4.6.3. 
To improve the image quality, the data were reprocessed with the EDSER algorithm enabled \citep{lietal2004} and
subsequently rebinned to 1/8th the native ACIS pixel size before convolving with a
FWHM=0.25$\arcsec$ Gaussian. 
Our spectral and variability analysis is performed on the event files.

\begin{figure*}[!t]
\begin{center}
\includegraphics[width=6.25in]{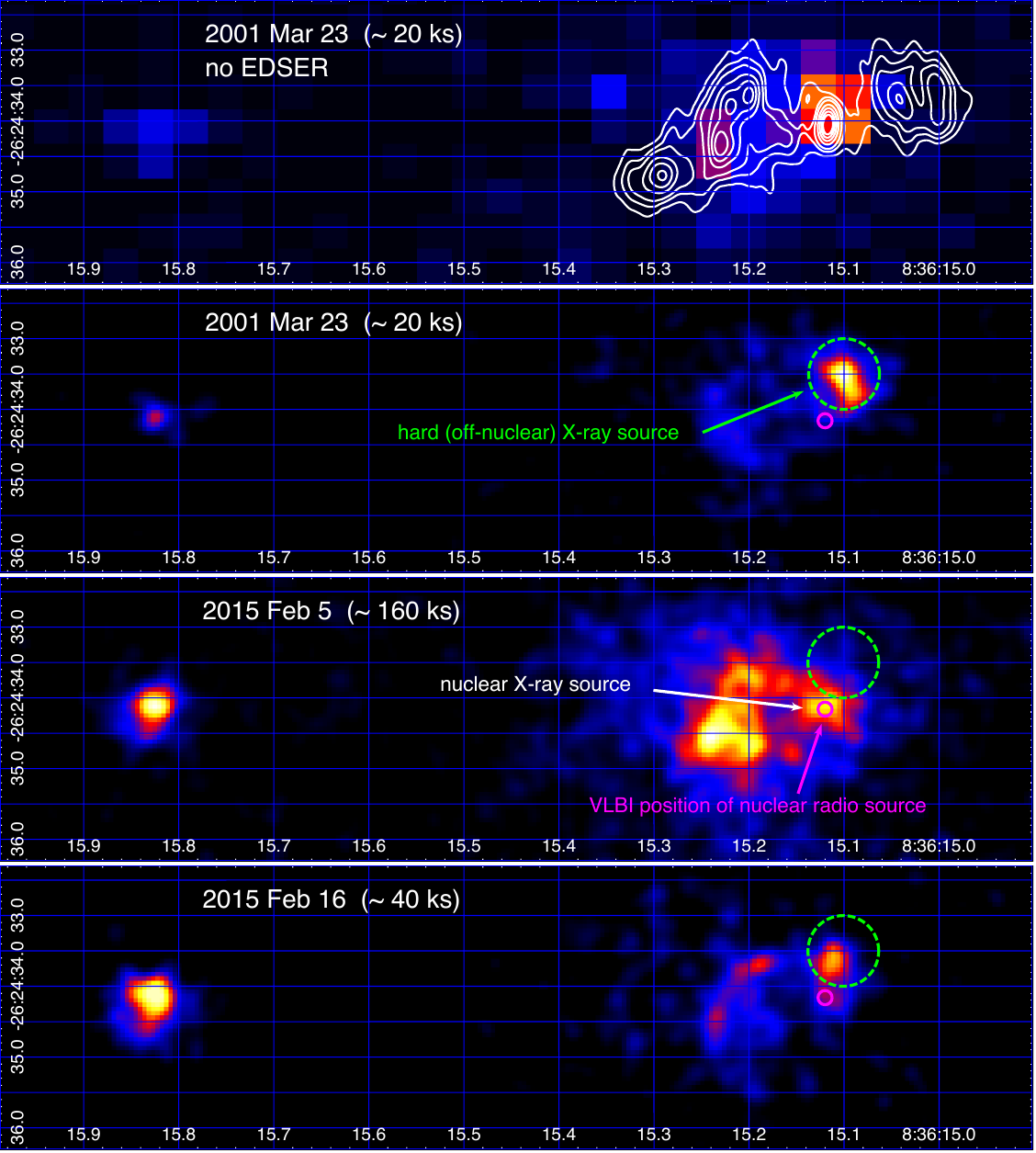}
\end{center}
\caption{{\it Chandra} observations showing the central region of Henize 2-10.  All images are in the 0.3-7 keV energy range but not on the same count rate scale.  The top panel shows the observation from 2001 used by \citet{reinesetal2011} without the EDSER algorithm enabled.  VLA contours from \citet{reinesetal2011} are overlaid to show the seeming match between the bright hard X-ray source and the nuclear radio source.  The bottom three panels show all observations using the EDSER algorithm.
The VLBI position and absolute positional uncertainty of the nuclear radio source from \citet{reinesdeller2012} is indicated by a magenta circle with $r=0\farcs1$.  The bright hard X-ray source previously identified in the 2001 observation (also seen in the 40 ks observation from 2015) is indicated by a green circle ($r=0\farcs5$) and is clearly offset from the nuclear radio source.  A newly revealed X-ray source visible in the 160 ks observation from 2015 is co-spatial with the nuclear radio source.    
\label{fig:3epochs}}
\end{figure*}

To improve the astrometry, we co-aligned the three {\it Chandra} observations and then tied the corrected images to the absolute reference frame defined by our radio observations, which is accurate to $\lesssim 0\farcs1$  \citep{reinesetal2011,reinesdeller2012}.  We first averaged the coordinates of a bright point source common to all three {\it Chandra} observations, yet outside the vicinity of the nuclear region.  This reference source is located $\sim9\farcs5$ east of the nuclear radio source and has a corrected (mean) position of RA=8:36:15.83, DEC=$-$26:24:34.1 (Figure \ref{fig:3epochs}).  We then registered the three {\it Chandra} observations by determining the relative offset between this mean position and the position of the reference source in an individual observation.  The required (RA, DEC) shifts in arcseconds are (0\farcs17 W, 0\farcs13 N), (0\farcs08 E, 0\farcs00 S), and (0\farcs08 E, 0\farcs13 S) for the 20~ks, 160~ks, and 40~ks observations, respectively.  There is no evidence for significant rotation between the different observations.  A comparison between our deep 160~ks observation and the VLA contours from \citet{reinesetal2011} indicates a close match between bright regions of X-ray and radio emission from recent star formation (inset Figure~\ref{fig:he210}), strongly suggesting the absolute astrometry of our corrected {\it Chandra} observations is accurate and requires no additional shift.  The absolute astrometric uncertainties of the final {\it Chandra} positions are estimated to be $0\farcs15$ in RA and $0\farcs13$ in DEC using the standard deviation of the three individual uncorrected measurements of the reference source.

\begin{figure*}[!t]
\begin{center}$
\begin{array}{cc}
\hspace{-0.3cm}
{\includegraphics[width=3.75in]{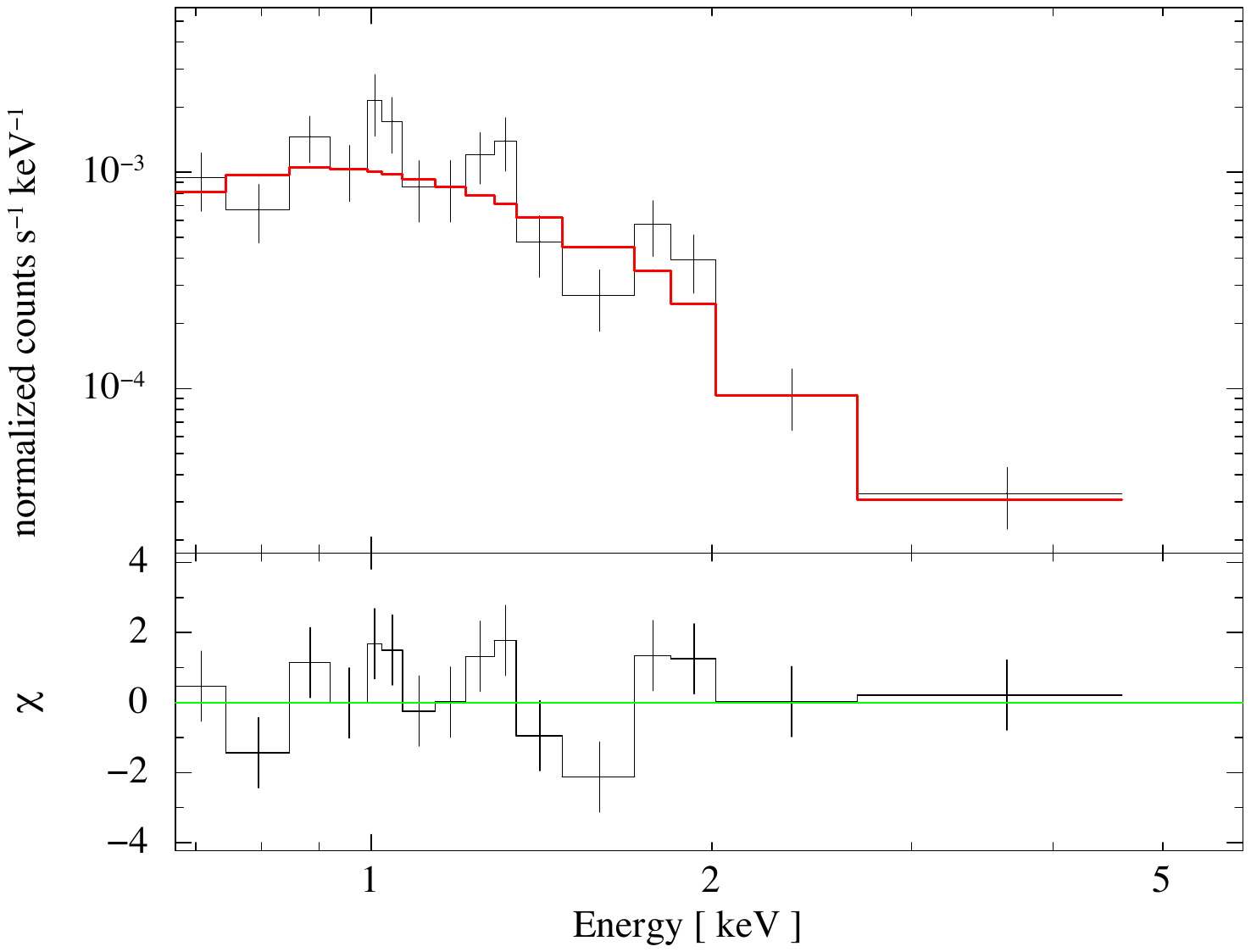}} &
\hspace{-0.5cm}
\vspace{-0.5cm}
{\includegraphics[width=3.75in]{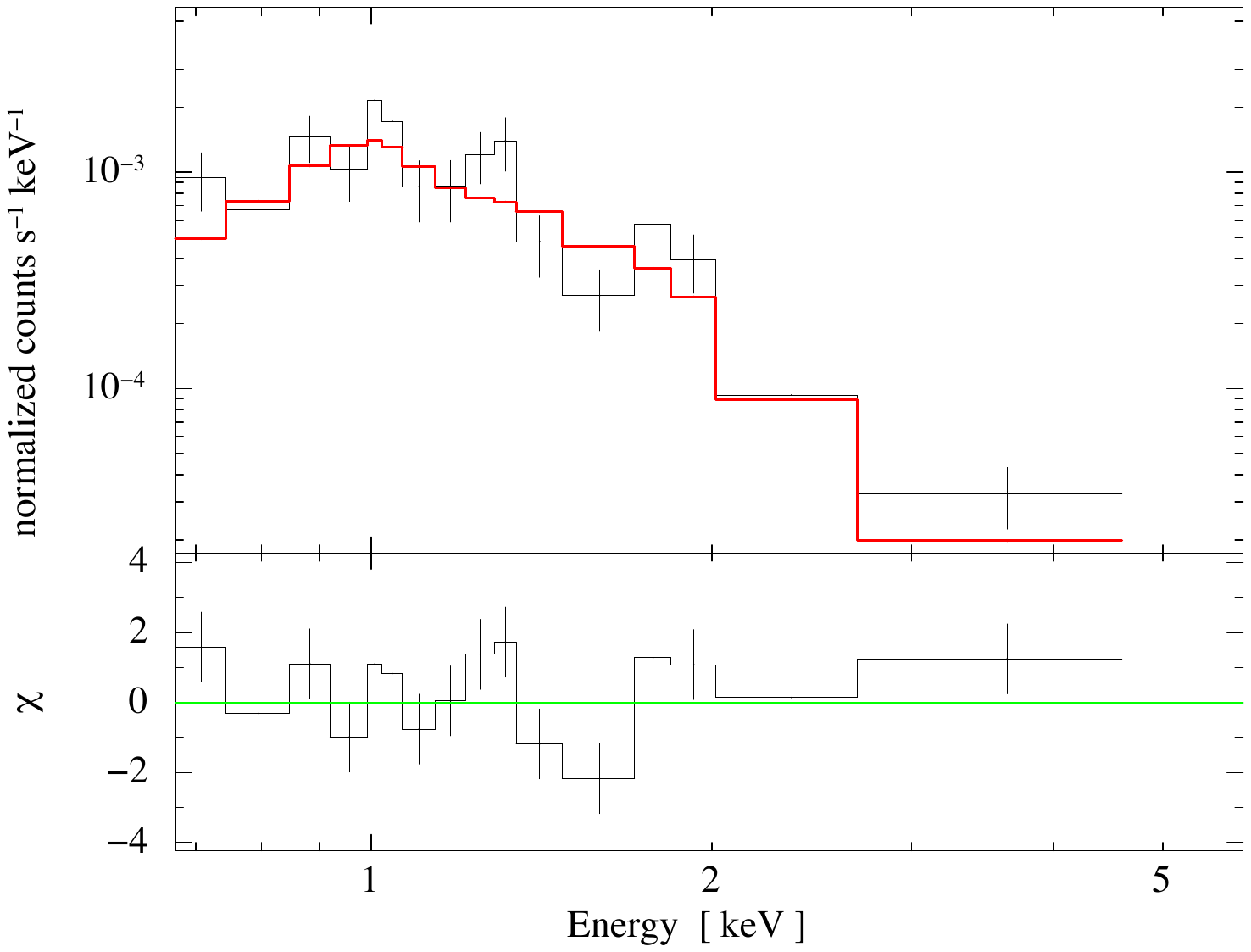}}
\end{array}$
\end{center}
\caption{X-ray spectrum of the nuclear source.  Best-fit power law
  model (left) and best-fit thermal plasma model (right), with residuals shown beneath.  
\label{fig:spec}}
\end{figure*}

\section{Analysis and Results}\label{sec:results}

\subsection{The Newly-Identifed Nuclear X-ray Source}

Our primary goal in this Letter is to examine the nuclear X-ray source in Henize 2-10.  From Figures \ref{fig:he210} and \ref{fig:3epochs}, we can now see that a previously unidentified X-ray source is spatially coincident with the central radio source, hereafter referred to as the nuclear X-ray source (visible in the 160 ks image at RA=8:36:15.12, DEC=$-$26:24:34.1), and that the bright X-ray emission identified in the 2001 observation \citep{ottetal2005,kobulnickymartin2010,reinesetal2011,whalenetal2015} is dominated by an X-ray source (RA=8:36:15.10, DEC=$-$26:24:33.5) distinct from the nuclear radio source.  The spatial offset between the bright off-nuclear X-ray source in the subpixel image from 2001 and the VLBI position is $\sim 0\farcs7$ ($\sim30$~pc projected). 
This offset is significantly larger than our astrometric uncertainties, and thus our conclusion that the hard X-ray source is distinct from the nuclear radio source is secure.  
This source is also highly variable -- it dominates the emission in 2001, is essentially absent in our new 160 ks observation, and returns in the 40 ks observation taken only 9 days later.  

The spectrum of the bright, off-nuclear variable source dominating the 2001 observation is poorly constrained, but consistent with a highly
absorbed power-law ($N_H \sim 7\times10^{22}~{\rm cm}^{-2}$, $\Gamma = 1.8$).
Assuming a common spectral shape in each of the three observations,
we measure unabsorbed 0.3-10.0 keV fluxes of $(5.09^{+1.68}_{-1.24}, 0.17^{+0.09}_{-0.07}$, and
$1.28^{+0.47}_{-0.38})\times10^{-13}$ erg s$^{-1}$ cm$^{-2}$ respectively (90\% confidence intervals), i.e., 
variability of a factor of $\sim 30$.
The large-amplitude variability
\citep[also see][]{whalenetal2015} and the large luminosity in the 2001
observation ($\ge 10^{39}~{\rm erg~s}^{-1}$) suggests it is an X-ray binary (XRB) containing a stellar-mass BH
primary.

\begin{deluxetable*}{lr}
\tabletypesize{\footnotesize}
\tablecaption{Spectral Fits to the Nuclear X-ray Source in Henize 2-10}
\tablewidth{0pt}
\startdata
\cutinhead{Power-Law Model}
$\chi^2/$dof \dotfill & 22.40/14 \\ [0.5ex]
$\Gamma$ \dotfill & 2.93$^{+0.36}_{-0.33}$  \\ [0.5ex]
Normalization ($10^{-6}$ photons cm$^{-2}$ s$^{-1}$  keV$^{-1}$ at 1 keV) \dotfill	&  4.30$\pm$0.66\\ [0.5ex]
$F_{\rm 0.3-10~keV}$ ($10^{-14}~{\rm erg~s^{-1}~cm^{-2}}$), absorbed \dotfill  & 1.30$^{+0.22}_{-0.17}$ \\ [0.5ex]
log $L_{\rm 0.3-10~keV}$ (${\rm erg~s^{-1}}$), unabsorbed \dotfill	& 38.10$^{+0.07}_{-0.06}$ \\ [0.5ex]
\cutinhead{Thermal Plasma Model (APEC)}
$\chi^2$/dof \dotfill & 22.64/13 \\ [0.5ex]
$kT$ (keV)  \dotfill & 1.06$^{+0.27}_{-0.19}$  \\ [0.5ex]
Z ($Z_{\sun}$)	 \dotfill	  & 0.06$^{+0.16}_{-0.06}$   \\ [0.5ex]
Normalization ($\frac{10^{-14}}{4\pi\{D_A(1+z)\}^2\int n_en_HdV}$) \dotfill	&  1.97$^{+0.87}_{-0.73}$\\ [0.5ex]
$F_{\rm 0.3-10~keV}$  ($10^{-14}~{\rm erg~s^{-1}~cm^{-2}}$), absorbed \dotfill &  0.89$^{+0.12}_{-0.28}$  \\ [0.5ex]
log $L_{\rm 0.3-10~keV}$ (${\rm erg~s^{-1}}$), unabsorbed \dotfill	& 37.94$^{+0.05}_{-0.16}$  \\ [0.5ex]
\enddata
\tablecomments{We adopt $N_{\rm H,Gal}=9.1\times10^{20}~{\rm cm}^{-2}$ due to the Milky Way, and $N_{\rm H}=1\times10^{20}~{\rm cm}^{-2}$ intrinsic to Henize 2-10.  The errors represent the 90\% confidence interval for one interesting parameter determined using the \texttt{error} command in \textsc{xspec}.} 
\label{tab:xspec}
\end{deluxetable*}

\subsection{X-ray Spectrum of the Nucleus}

The fortuitous disappearance of the highly variable off-nuclear X-ray source in our 160 ks
observation enables us to extract a relatively clean X-ray spectrum of the nuclear source.
We extracted the spectrum in the 0.3-7.0~keV range from a circular aperture
with a radius of $0\farcs5$, correcting for the small aperture. The background
was estimated from a source-free annular region extending from 20-25$\arcsec$ centered on
the nuclear source.  We note that the external background contribution is negligible at
the position of the nucleus; however, we cannot reliably separate the point source from
any local background within the source extraction region. We obtained a total of 183 net counts. The spectrum was grouped to
have SNR=3 per bin, and spectral fits to the background subtracted spectrum were carried
out within {\footnotesize XSPEC} 12.8.2q \citep{arnaud1996} using the chi statistic and standard Gaussian weighting. Galactic foreground absorption
was held fixed at $N_{\rm H,Gal} = 9.1 \times
10^{20}$ cm$^{-2}$ \citep{kalberlaetal2005} and the internal absorption
was found to be negligible.
We used the \texttt{phabs} absorption model, with abundances and cross-sections adopted from  \citet{asplundetal2009} 
and \citet{balucinskachurchetal1992}, respectively.

The source spectrum is well characterized by either an absorbed
power-law model with photon index $\Gamma \sim 2.9$, or a thermal plasma model with $kT
\sim 1.1$ keV (Figure \ref{fig:spec}, Table \ref{tab:xspec}). Both models give a consistent intrinsic luminosity
of $L_{\rm 0.3-10~keV} \sim 10^{38}$ erg s$^{-1}$.

\subsection{X-ray Light Curve of the Nucleus}

We also examine the temporal behavior of the nuclear X-ray emission during the 160 ks
observation. Utilizing the same extraction regions as used for the
spectral analysis, a background subtracted light curve was created using
\textsc{dmextract}. 
As the count rate is low, the light curve was extracted at 5 ks resolution and Gehrels errors are assumed \citep{gehrels1986}.

\begin{figure*}[!t]
\begin{center}$
\begin{array}{cc}
\hspace{-0.2cm}
{\includegraphics[width=3.7in]{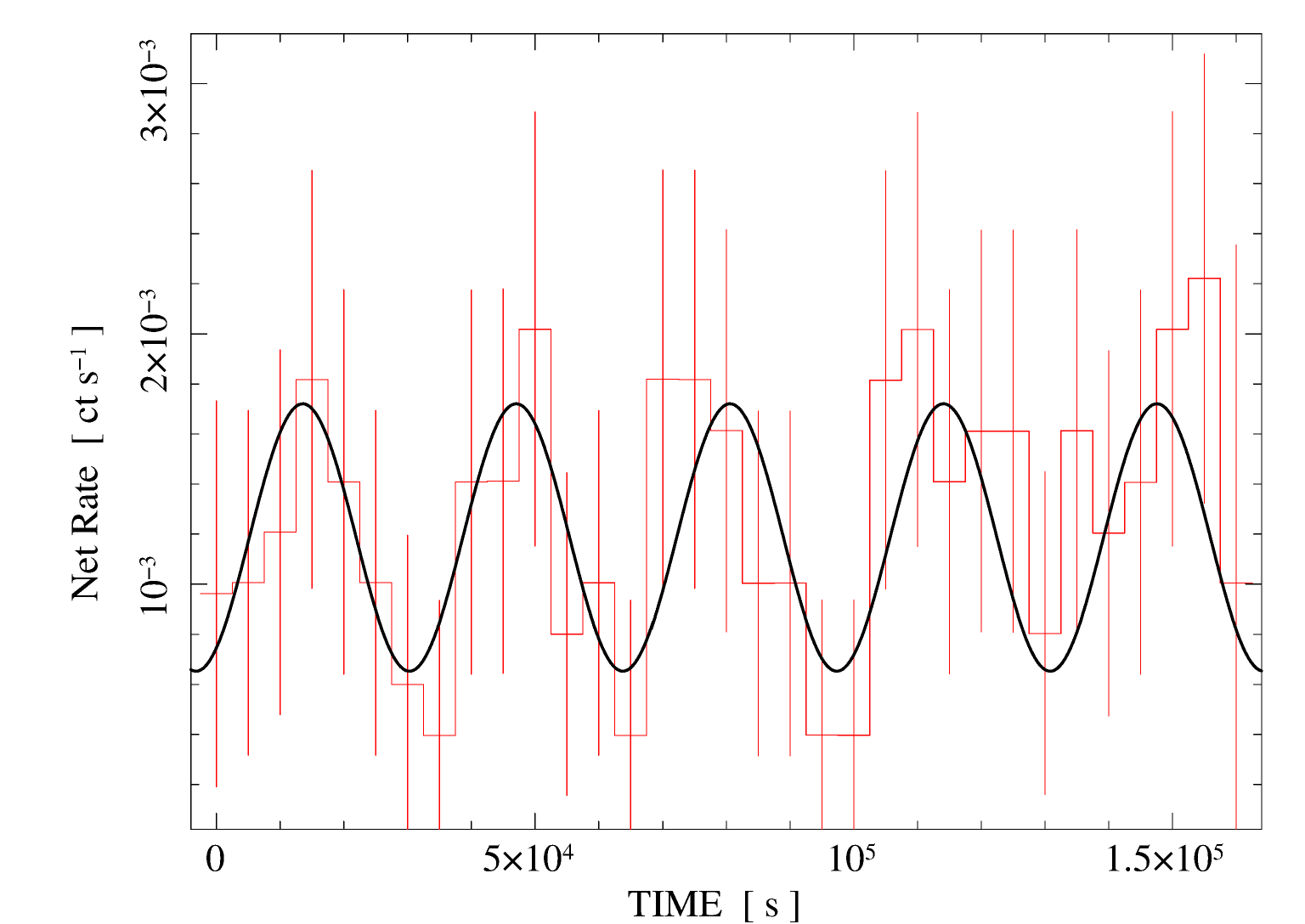}} &
\hspace{-0.5cm}
\vspace{-0.5cm}
{\includegraphics[width=3.7in]{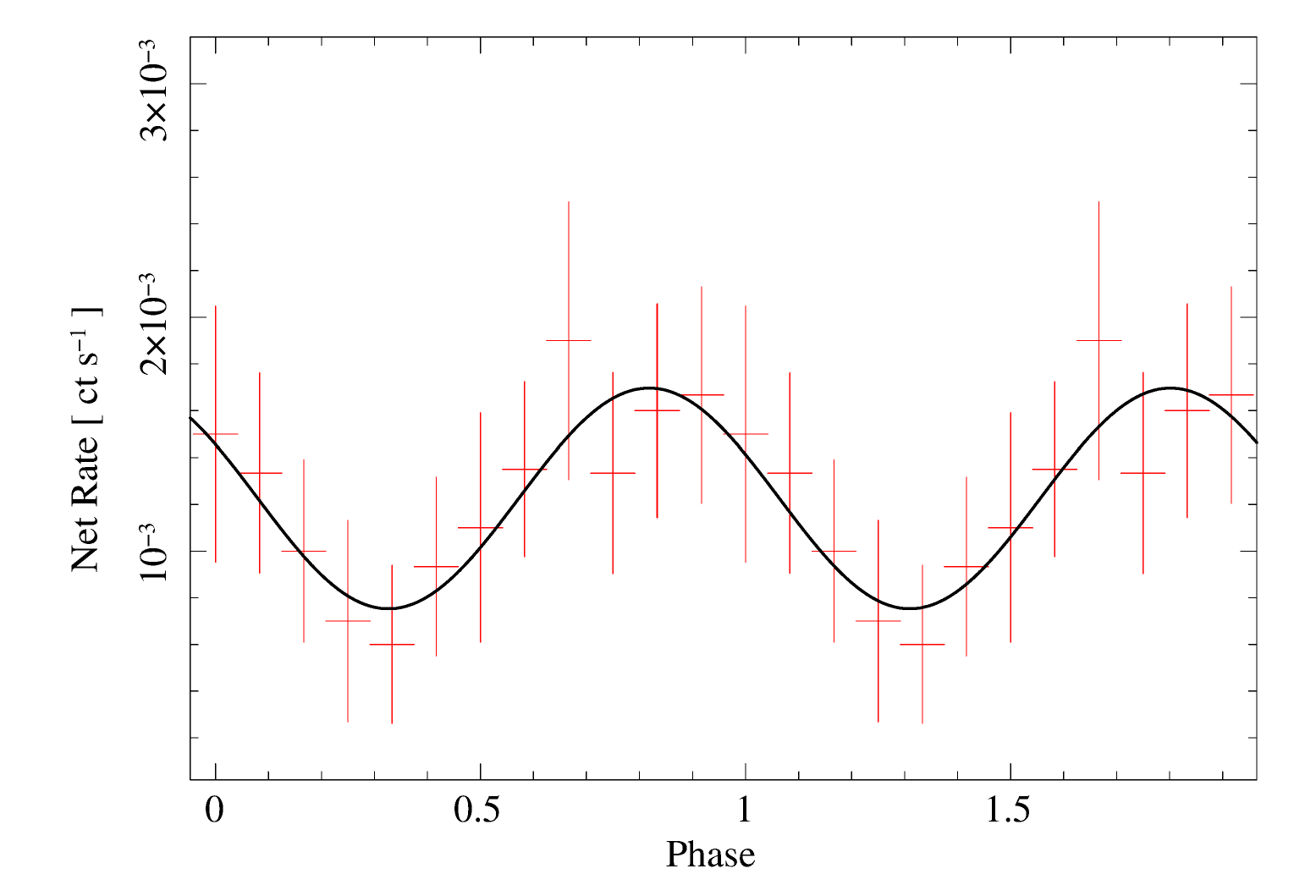}}
\end{array}$
\end{center}
\caption{Oscillating X-ray emission from the nuclear source.  Left: Light curve binned
  at 5 ks resolution with the best fit sine model overplotted. Right: Light curve
  folded on the best fit period of $P = 33.5$~ks and rebinned to 12 bins per
  cycle. Two cycles are displayed for clarity.
\label{fig:lightcurve}}
\end{figure*}

From Figure \ref{fig:lightcurve}, we can see that the light curve exhibits clear variability (a factor of $\sim 2\times$) and 
the X-ray emission appears to oscillate within
the 160 ks exposure.  A model consisting of a constant plus a sine wave provides an
excellent characterization of the light curve ($\chi^2/\nu = 9.2/29$) and reveals a best
fit period of $P = 33.5\pm2.6$~ks (90\% confidence level).   In the right panel of Figure \ref{fig:lightcurve}, we plot the
resulting light curve when folded on the detected period of $P$ = 33.5 ks (9.3 hrs).  The amplitude of the sine model is $(4.71 \pm 1.22) \times 10^{-4}$ counts sec$^{-1}$, and thus is measured to $3.9\sigma$. 

If instead the light curve is modeled as a constant, we obtain a best fit of $\chi^2/\nu =
18.4/32$. An F-test was used to determine that the sine model is a superior description of
the data at the 99.9866\% confidence level ($3.8\sigma$).
This simple statistical test is dependent
on the binning of the light curve, e.g., a binning of 3/10 ks resolution favors a sine wave
over a constant model at the 95.7\%/99.3\% level with $P = 33.4\pm2.2~{\rm ks}, 33.8\pm1.9$~ks
respectively. However, the best fit period is robustly determined irrespective of the actual
binning.  

The power-spectrum also hints at the presence of a periodic signal.
Utilizing the entire observation at the native temporal resolution ($\Delta$t = 3.14104 s)
reveals the presence of a low significance peak ($\lesssim 2\sigma$) at a frequency $f \sim 3\times
10^{-5}~{\rm Hz}$ ($P \sim 33$~ks) on top of a white noise background. 
If the signal is in fact periodic, the detected peak in the power-spectrum is likely of low statistical
significance due to the small number of cycles sampled in our 160 ks observation
($\lesssim 5$).  However, we cannot rule out the possibility that the 
seemingly periodic signal could be produced by
stochastic variability, which can sometimes mimic intervals of periodicity (e.g., ``red noise,"
\citealt{vaughanuttley2006}; \citealt{vaughanetal2016}).

The oscillating signal is uniquely coincident with the nuclear source
as identified in our sub-pixel spatial analysis. We have examined light curves from the
bright region 2$\arcsec$ to the east and find no evidence for any significant periodic
signal at this location only 3 ACIS-S pixels from the nucleus. Likewise, analysis
of the light curves of the remaining point sources on the ACIS-S3 detector reveals no
evidence for periodicity. Neither do we observe significant background flaring or the
presence of periodic variability in the background signal on the ACIS-S3 detector during
this observation.  A period of $\sim$~33 ks should also not be due to aspect dithering. 

\section{Discussion}\label{sec:discussion}

We have shown that a previously unidentified {\it Chandra} X-ray point
source is spatially coincident ($\lesssim 0\farcs1$, $\lesssim 5$ pc projected) with the non-thermal compact
nuclear radio source in Henize 2-10 \citep{reinesdeller2012}. Using three separate {\it Chandra} observations and
improved image processing, we have
determined that the bright hard X-ray source that dominated the original 2001 observation
\citep{ottetal2005,kobulnickymartin2010,reinesetal2011,whalenetal2015} is distinct and
spatially offset from the nuclear radio source. 
The previously detected hard X-ray source is highly variable with
properties expected for a stellar-mass BH XRB.
Fortunately, this source was absent in our new 160~ks observation,
which enabled us to extract and analyze a clean spectrum and light curve of the newly revealed nuclear
X-ray source.

Our new results from {\it Chandra} support a massive BH origin for
the nuclear source in Henize 2-10 \citep{reinesetal2011}.  
Given a recent estimate for the total stellar mass of
Henize 2-10 ($M_\star \sim 10^{10} M_\odot$; \citealt{nguyenetal2014}), we expect a
nuclear BH with a mass\footnote{Coincidentally, this is
  approximately the same mass estimated by \citet{reinesetal2011} using the BH fundamental
  plane \citep{merlonietal2003} and the luminosity of the bright variable source that
  dominated the 2001 {\it Chandra} observation.  Using the luminosity of the nuclear
  source in our new observation ($L_{\rm X} \sim 10^{38}$ erg s$^{-1}$,
  when the transient source was ``off") and a 5 GHz radio luminosity of $L_{\rm R} \sim 4
  \times10^{35}$ erg s$^{-1}$ \citep{reinesdeller2012}, the fundamental plane gives
  log($M_{\rm BH}/M_\odot) \sim 7 \pm 1$, consistent with the estimate based on the
  scaling between $M_{\rm BH}$ and $M_\star$ from \citet{reinesvolonteri2015}.} of $M_{\rm BH} \sim 3 \times 10^6 M_\odot$  \citep{reinesvolonteri2015}, although the uncertainty is at
least a factor of a few.  
 The luminosity of the nuclear X-ray source is $L_{0.3-10~{\rm keV}} \sim 10^{38}$ erg
s$^{-1}$.  This implies an
Eddington ratio of $\sim$$10^{-6}$, assuming an X-ray to bolometric
correction of $\sim$10 \citep[e.g.,][]{vasudevanfabian2009}.  The X-ray spectrum is soft
and can be well-fit by either a thermal plasma model with $kT \sim 1.1$ keV or a power-law
model with $\Gamma \sim 2.9$, similar to Sagittarius A$^{*}$ at the center of the Milky
Way \citep{baganoffetal2003} and other massive BHs accreting at very low Eddington ratios
\citep[e.g.,][]{constantinetal2009}.  The presence of a spatially coincident non-thermal
radio source, with a physical size of $\lesssim 1~{\rm pc} \times 3~{\rm pc}$
\citep{reinesdeller2012}, also strongly suggests a massive BH.  Furthermore, the nuclear source in
Henize 2-10 falls along the correlation between nuclear radio and X-ray luminosity for low-luminosity radio
galaxies (within the $1\sigma$ scatter) found by \citet{panessaetal2007}.  

Our temporal analysis of the nuclear X-ray emission reveals clear variability and a potential $\sim9$-hr periodicity.  
The oscillatory signal is apparent by eye and significant when fitting the light curve.  However, the X-ray periodicity
is not yet significant when considering the power spectrum, and at present we cannot distinguish between a true
periodic signal and random fluctuations in brightness mimicking a periodic signal during our 160~ks observation 
(e.g., red noise, see \citealt{vaughanuttley2006}; \citealt{vaughanetal2016}).  Nevertheless, we
discuss possible origins for a $\sim9$-hr X-ray periodicity below.

The most likely origin is 
a low frequency quasi-periodic oscillation (LFQPO; \citealt{remillardmcclintock2006}). 
QPOs are generally thought to arise from instabilities in the accretion flow \citep[e.g.,][]{taggerpellat1999} or geometric oscillations \citep[e.g.,][]{chakrabartimolteni1993}.
It has been proposed that LFQPOs may be due to the orbital precession of non-equitorial particles in the dragged
spacetime around a spinning BH (i.e., Lense-Thirring precession; \citealt{stellarvietri1998}).
Given the timescale of the apparent periodicity and
the low luminosity, current observational constraints on such 
variability are limited, with only a single claimed detection of a LFQPO from
a massive BH that is accreting at a relatively high rate \citep[$\sim 0.1~L_{\rm Edd}$,][]{linetal2013}. 
 
We also considered a high frequency QPO (HFQPO) as an origin for the X-ray periodicity,
however the observed luminosity and frequency argue against this.
The known stellar and massive BH QPO detections that are
consistent with a HFQPO origin are all from sources known to be accreting at close to the
Eddington limit (e.g., \citealt{remillardmcclintock2006}; \citealt{gierlinskietal2008},
\citealt{alstonetal2014}; \citealt{reisetal2012}; 
\citealt{panetal2016}; \citealt{alstonetal2015}). 
HFQPOs are thought to originate in the
inner regions of the accretion flow in the immediate vicinity of the BH.  The
observed relation between QPO frequency and BH mass (e.g.,
\citealt{remillardmcclintock2006,zhouetal2015,panetal2016}) would predict a mass of $\sim
10^8~M_{\sun}$ for a frequency of $f \sim 3\times10^{-5}$~Hz, more than an order of magnitude
greater than the expected mass of the BH in Henize 2-10.

A final intriguing possibility is orbital variability related
to a massive BH binary, where at least one of the BHs is actively accreting, albeit at a
low level \citep{roedigetal2012,roedigetal2014,sesana2013}. In this case the periodicity would
correspond to the serendipitous electromagnetic discovery of a massive BH binary with $\lesssim
5$ years until merger, for an assumed total system mass of $\sim 10^6~M_{\sun}$
\citep{sesana2013}. Although there is evidence that Henize 2-10 has experienced a merger in the
recent past \citep{kobulnickyetal1995}, and thus the presence of two massive BHs is possible, we regard
this possibility as unlikely given the extremely short timescale until coalescence
implied by the observed frequency.

While we consider a weakly accreting massive BH the most likely origin for the nuclear
X-ray/radio source in Henize 2-10,
we nonetheless revisit alternative explanations including a stellar-mass XRB and/or a
young supernova remnant (SNR; also see the Supplementary Information in
\citealt{reinesetal2011} as well as the discussion in \citealt{reinesdeller2012}).
Neither an XRB nor a SNR alone can account for \textit{both} the X-ray and radio
properties of the nuclear source.  The radio emission is simply too luminous to be
produced by an XRB, especially one in the soft thermal state \citep{fenderetal2009} as would be indicated by the observed spectrum, and the
X-ray variability is incompatible with a SNR.  
Our constraints on the positions of the nuclear radio and X-ray sources, however, strongly suggest 
a common/related source. 
In principle, we can imagine a scenario
in which the X-rays originate from an accreting stellar-mass XRB residing within the radio-emitting remnant of the supernova that created the compact object.
Given the size and luminosity of the radio source, a SNR would likely be only decades old \citep{reinesdeller2012,fenechetal2010} and therefore the nuclear X-ray source would be the new record-holder for the youngest XRB known by a wide margin \citep[e.g., Circinus X-1 has an age $t < 4600$ yr;][]{heinzeetal2013}.  While we cannot definitively rule out this scenario, we consider it somewhat contrived.  Moreover, the radio/X-ray source is at the center of the galaxy (the natural place for a massive BH) and there is no star cluster or recent star formation at the location of the source as would be expected for a young XRB/SNR origin \citep{reinesetal2011}.  

Finally, our study demonstrates the value of X-ray imaging at sub-arcsecond scales and emphasizes the
need for a high-resolution next generation X-ray observatory.  Our detection of the first potential LFQPO from a low-luminosity massive BH would not have been possible with any existing X-ray observatory other than {\it Chandra}.  Future studies of this class of objects (e.g., with {\it ATHENA} and the X-ray Surveyor) would open a new window to study low luminosity accretion flows, which are the dominant mode of BH accretion on
cosmological scales.
          
\acknowledgements

We thank Richard Plotkin for useful discussions and the anonymous referee for reviewing our work.
AER is grateful for support from NASA through Hubble Fellowship grant HST-HF2-51347.001-A
awarded by the Space Telescope Science Institute, which is operated by the Association of
Universities for Research in Astronomy, Inc., for NASA, under contract NAS 5-26555.
Support for this work was also provided by NASA through Chandra Award Number GO4-15098A
issued by the Chandra X-ray Observatory Center, which is operated by the Smithsonian
Astrophysical Observatory for and on behalf of the NASA under contract NAS8-03060.  GRS
acknowledges support by an NSERC Discovery Grant.


\begin{thebibliography}{}
\expandafter\ifx\csname natexlab\endcsname\relax\def\natexlab#1{#1}\fi

\bibitem[{{Alston} {et~al.}(2014){Alston}, {Markevi{\v c}i{\= u}t{\.e}},
  {Kara}, {Fabian}, \& {Middleton}}]{alstonetal2014}
{Alston}, W.~N., {Markevi{\v c}i{\= u}t{\.e}}, J., {Kara}, E., {Fabian}, A.~C.,
  \& {Middleton}, M. 2014, \mnras, 445, L16

\bibitem[{{Alston} {et~al.}(2015){Alston}, {Parker}, {Markevi{\v c}i{\=
  u}t{\.e}}, {Fabian}, {Middleton}, {Lohfink}, {Kara}, \&
  {Pinto}}]{alstonetal2015}
{Alston}, W.~N., {Parker}, M.~L., {Markevi{\v c}i{\= u}t{\.e}}, J., {et~al.}
  2015, \mnras, 449, 467

\bibitem[{{Arca-Sedda} {et~al.}(2015){Arca-Sedda}, {Capuzzo-Dolcetta},
  {Antonini}, \& {Seth}}]{arcaseddaetal2015}
{Arca-Sedda}, M., {Capuzzo-Dolcetta}, R., {Antonini}, F., \& {Seth}, A. 2015,
  \apj, 806, 220

\bibitem[{{Arnaud}(1996)}]{arnaud1996}
{Arnaud}, K.~A. 1996, in Astronomical Society of the Pacific Conference Series,
  Vol. 101, Astronomical Data Analysis Software and Systems V, ed. G.~H.
  {Jacoby} \& J.~{Barnes}, 17

\bibitem[{{Asplund} {et~al.}(2009){Asplund}, {Grevesse}, {Sauval}, \&
  {Scott}}]{asplundetal2009}
{Asplund}, M., {Grevesse}, N., {Sauval}, A.~J., \& {Scott}, P. 2009, \araa, 47,
  481

\bibitem[{{Baganoff} {et~al.}(2003){Baganoff}, {Maeda}, {Morris}, {Bautz},
  {Brandt}, {Cui}, {Doty}, {Feigelson}, {Garmire}, {Pravdo}, {Ricker}, \&
  {Townsley}}]{baganoffetal2003}
{Baganoff}, F.~K., {Maeda}, Y., {Morris}, M., {et~al.} 2003, \apj, 591, 891

\bibitem[{{Baldassare} {et~al.}(2015){Baldassare}, {Reines}, {Gallo}, \&
  {Greene}}]{baldassareetal2015}
{Baldassare}, V.~F., {Reines}, A.~E., {Gallo}, E., \& {Greene}, J.~E. 2015,
  \apjl, 809, L14

\bibitem[Baldassare et al.(2016)]{baldassareetal2016} Baldassare, V.~F., Reines, A.~E., Gallo, E., \& Greene, J.~E.\ 2016, arXiv:1609.07148 

\bibitem[{{Balucinska-Church} \& {McCammon}(1992)}]{balucinskachurchetal1992}
{Balucinska-Church}, M., \& {McCammon}, D. 1992, \apj, 400, 699

\bibitem[{{Chakrabarti} \& {Molteni}(1993)}]{chakrabartimolteni1993}
{Chakrabarti}, S.~K., \& {Molteni}, D. 1993, \apj, 417, 671


\bibitem[{{Constantin} {et~al.}(2009){Constantin}, {Green}, {Aldcroft}, {Kim},
  {Haggard}, {Barkhouse}, \& {Anderson}}]{constantinetal2009}
{Constantin}, A., {Green}, P., {Aldcroft}, T., {et~al.} 2009, \apj, 705, 1336

\bibitem[{{Fender} {et~al.}(2009){Fender}, {Homan}, \&
  {Belloni}}]{fenderetal2009}
{Fender}, R.~P., {Homan}, J., \& {Belloni}, T.~M. 2009, \mnras, 396, 1370

\bibitem[{{Fenech} {et~al.}(2010){Fenech}, {Beswick}, {Muxlow}, {Pedlar}, \&
  {Argo}}]{fenechetal2010}
{Fenech}, D., {Beswick}, R., {Muxlow}, T.~W.~B., {Pedlar}, A., \& {Argo}, M.~K.
  2010, \mnras, 408, 607

\bibitem[{{Fruscione} {et~al.}(2006){Fruscione}, {McDowell}, {Allen},
  {Brickhouse}, {Burke}, {Davis}, {Durham}, {Elvis}, {Galle}, {Harris},
  {Huenemoerder}, {Houck}, {Ishibashi}, {Karovska}, {Nicastro}, {Noble},
  {Nowak}, {Primini}, {Siemiginowska}, {Smith}, \& {Wise}}]{fruscioneetal2006}
{Fruscione}, A., {McDowell}, J.~C., {Allen}, G.~E., {et~al.} 2006, in Society
  of Photo-Optical Instrumentation Engineers (SPIE) Conference Series, Vol.
  6270, Society of Photo-Optical Instrumentation Engineers (SPIE) Conference
  Series

\bibitem[{{Gehrels}(1986)}]{gehrels1986}
{Gehrels}, N. 1986, \apj, 303, 336

\bibitem[{{Gierli{\'n}ski} {et~al.}(2008){Gierli{\'n}ski}, {Middleton}, {Ward},
  \& {Done}}]{gierlinskietal2008}
{Gierli{\'n}ski}, M., {Middleton}, M., {Ward}, M., \& {Done}, C. 2008, \nat,
  455, 369

\bibitem[Hainline et al.(2016)]{hainlineetal2016} Hainline, K.~N., Reines, A.~E., Greene, J.~E., \& Stern, D.\ 2016, arXiv:1609.06721 

\bibitem[{{Heinz} {et~al.}(2013){Heinz}, {Sell}, {Fender}, {Jonker}, {Brandt},
  {Calvelo-Santos}, {Tzioumis}, {Nowak}, {Schulz}, {Wijnands}, \& {van der
  Klis}}]{heinzeetal2013}
{Heinz}, S., {Sell}, P., {Fender}, R.~P., {et~al.} 2013, \apj, 779, 171

\bibitem[{{Johnson} \& {Kobulnicky}(2003)}]{johnsonkobulnicky2003}
{Johnson}, K.~E., \& {Kobulnicky}, H.~A. 2003, \apj, 597, 923

\bibitem[{{Johnson} {et~al.}(2000){Johnson}, {Leitherer}, {Vacca}, \&
  {Conti}}]{johnsonetal2000}
{Johnson}, K.~E., {Leitherer}, C., {Vacca}, W.~D., \& {Conti}, P.~S. 2000, \aj,
  120, 1273

\bibitem[{{Kalberla} {et~al.}(2005){Kalberla}, {Burton}, {Hartmann}, {Arnal},
  {Bajaja}, {Morras}, \& {P{\"o}ppel}}]{kalberlaetal2005}
{Kalberla}, P.~M.~W., {Burton}, W.~B., {Hartmann}, D., {et~al.} 2005, \aap,
  440, 775

\bibitem[{{Kobulnicky} {et~al.}(1995){Kobulnicky}, {Dickey}, {Sargent}, {Hogg},
  \& {Conti}}]{kobulnickyetal1995}
{Kobulnicky}, H.~A., {Dickey}, J.~M., {Sargent}, A.~I., {Hogg}, D.~E., \&
  {Conti}, P.~S. 1995, \aj, 110, 116

\bibitem[{{Kobulnicky} \& {Johnson}(1999)}]{kobulnickyjohnson1999}
{Kobulnicky}, H.~A., \& {Johnson}, K.~E. 1999, \apj, 527, 154

\bibitem[{{Kobulnicky} \& {Martin}(2010)}]{kobulnickymartin2010}
{Kobulnicky}, H.~A., \& {Martin}, C.~L. 2010, \apj, 718, 724

\bibitem[{{Lemons} {et~al.}(2015){Lemons}, {Reines}, {Plotkin}, {Gallo}, \&
  {Greene}}]{lemonsetal2015}
{Lemons}, S.~M., {Reines}, A.~E., {Plotkin}, R.~M., {Gallo}, E., \& {Greene},
  J.~E. 2015, \apj, 805, 12

\bibitem[{{Li} {et~al.}(2004){Li}, {Kastner}, {Prigozhin}, {Schulz},
  {Feigelson}, \& {Getman}}]{lietal2004}
{Li}, J., {Kastner}, J.~H., {Prigozhin}, G.~Y., {et~al.} 2004, \apj, 610, 1204

\bibitem[{{Lin} {et~al.}(2013){Lin}, {Irwin}, {Godet}, {Webb}, \&
  {Barret}}]{linetal2013}
{Lin}, D., {Irwin}, J.~A., {Godet}, O., {Webb}, N.~A., \& {Barret}, D. 2013,
  \apjl, 776, L10

\bibitem[{{Merloni} {et~al.}(2003){Merloni}, {Heinz}, \& {di
  Matteo}}]{merlonietal2003}
{Merloni}, A., {Heinz}, S., \& {di Matteo}, T. 2003, \mnras, 345, 1057

\bibitem[{{Nguyen} {et~al.}(2014){Nguyen}, {Seth}, {Reines}, {den Brok},
  {Sand}, \& {McLeod}}]{nguyenetal2014}
{Nguyen}, D.~D., {Seth}, A.~C., {Reines}, A.~E., {et~al.} 2014, \apj, 794, 34

\bibitem[{{Ott} {et~al.}(2005){Ott}, {Walter}, \& {Brinks}}]{ottetal2005}
{Ott}, J., {Walter}, F., \& {Brinks}, E. 2005, \mnras, 358, 1423

\bibitem[{{Pan} {et~al.}(2016){Pan}, {Yuan}, {Yao}, {Zhou}, {Liu}, {Zhou}, \&
  {Zhang}}]{panetal2016}
{Pan}, H.-W., {Yuan}, W., {Yao}, S., {et~al.} 2016, \apjl, 819, L19

\bibitem[{{Panessa} {et~al.}(2007){Panessa}, {Barcons}, {Bassani}, {Cappi},
  {Carrera}, {Ho}, \& {Pellegrini}}]{panessaetal2007}
{Panessa}, F., {Barcons}, X., {Bassani}, L., {et~al.} 2007, \aap, 467, 519

\bibitem[Reines \& Comastri(2016)]{reinescomastri2016} Reines, A., \& Comastri, A.\ 2016, arXiv:1609.03562 

\bibitem[{{Reines} \& {Deller}(2012)}]{reinesdeller2012}
{Reines}, A.~E., \& {Deller}, A.~T. 2012, \apjl, 750, L24

\bibitem[{{Reines} {et~al.}(2013){Reines}, {Greene}, \&
  {Geha}}]{reinesetal2013}
{Reines}, A.~E., {Greene}, J.~E., \& {Geha}, M. 2013, \apj, 775, 116

\bibitem[{{Reines} {et~al.}(2014){Reines}, {Plotkin}, {Russell}, {Mezcua},
  {Condon}, {Sivakoff}, \& {Johnson}}]{reinesetal2014}
{Reines}, A.~E., {Plotkin}, R.~M., {Russell}, T.~D., {et~al.} 2014, \apjl, 787,
  L30

\bibitem[{{Reines} {et~al.}(2011){Reines}, {Sivakoff}, {Johnson}, \&
  {Brogan}}]{reinesetal2011}
{Reines}, A.~E., {Sivakoff}, G.~R., {Johnson}, K.~E., \& {Brogan}, C.~L. 2011,
  \nat, 470, 66

\bibitem[{{Reines} \& {Volonteri}(2015)}]{reinesvolonteri2015}
{Reines}, A.~E., \& {Volonteri}, M. 2015, \apj, 813, 82

\bibitem[{{Reis} {et~al.}(2012){Reis}, {Miller}, {Reynolds}, {G{\"u}ltekin},
  {Maitra}, {King}, \& {Strohmayer}}]{reisetal2012}
{Reis}, R.~C., {Miller}, J.~M., {Reynolds}, M.~T., {et~al.} 2012, Science, 337,
  949

\bibitem[{{Remillard} \& {McClintock}(2006)}]{remillardmcclintock2006}
{Remillard}, R.~A., \& {McClintock}, J.~E. 2006, \araa, 44, 49

\bibitem[{{Roedig} {et~al.}(2014){Roedig}, {Krolik}, \&
  {Miller}}]{roedigetal2014}
{Roedig}, C., {Krolik}, J.~H., \& {Miller}, M.~C. 2014, \apj, 785, 115

\bibitem[{{Roedig} {et~al.}(2012){Roedig}, {Sesana}, {Dotti}, {Cuadra},
  {Amaro-Seoane}, \& {Haardt}}]{roedigetal2012}
{Roedig}, C., {Sesana}, A., {Dotti}, M., {et~al.} 2012, \aap, 545, A127

\bibitem[{{Sesana}(2013)}]{sesana2013}
{Sesana}, A. 2013, Classical and Quantum Gravity, 30, 244009

\bibitem[{{Stella} \& {Vietri}(1998)}]{stellarvietri1998}
{Stella}, L., \& {Vietri}, M. 1998, \apjl, 492, L59

\bibitem[{{Tagger} \& {Pellat}(1999)}]{taggerpellat1999}
{Tagger}, M., \& {Pellat}, R. 1999, \aap, 349, 1003

\bibitem[{{Vasudevan} \& {Fabian}(2009)}]{vasudevanfabian2009}
{Vasudevan}, R.~V., \& {Fabian}, A.~C. 2009, \mnras, 392, 1124

\bibitem[{{Vaughan} \& {Uttley}(2006)}]{vaughanuttley2006}
{Vaughan}, S., \& {Uttley}, P. 2006, Advances in Space Research, 38, 1405

\bibitem[{{Vaughan} {et~al.}(2016){Vaughan}, {Uttley}, {Markowitz},
  {Huppenkothen}, {Middleton}, {Alston}, {Scargle}, \&
  {Farr}}]{vaughanetal2016}
{Vaughan}, S., {Uttley}, P., {Markowitz}, A.~G., {et~al.} 2016, \mnras, 461,
  3145

\bibitem[{{Whalen} {et~al.}(2015){Whalen}, {Hickox}, {Reines}, {Greene},
  {Sivakoff}, {Johnson}, {Alexander}, \& {Goulding}}]{whalenetal2015}
{Whalen}, T.~J., {Hickox}, R.~C., {Reines}, A.~E., {et~al.} 2015, \apj, 806, 37

\bibitem[{{Zhou} {et~al.}(2015){Zhou}, {Yuan}, {Pan}, \& {Liu}}]{zhouetal2015}
{Zhou}, X.-L., {Yuan}, W., {Pan}, H.-W., \& {Liu}, Z. 2015, \apjl, 798, L5

\end{thebibliography}
\end{document}